\documentclass[11pt]{article}

\usepackage[preprint]{acl}

\usepackage{times}
\usepackage{latexsym}

\usepackage[T1]{fontenc}

\usepackage[utf8]{inputenc}

\usepackage{microtype}

\usepackage{inconsolata}

\usepackage{graphicx}
\usepackage{multirow}
\usepackage{booktabs}
\usepackage{amsmath}
\usepackage{tabularx}
\usepackage{amsfonts}
\title{Reward-Guided Autoregressive Graph Generation for Efficient Multi-Agent Communication Topology Design}

\author{
  Poomphob Suwannapichat$^{1}$ \quad
  Boonyarit Changaival$^{2}$ \quad
  Caesar Wu$^{1}$ \quad
  Pascal Bouvry$^{1}$ \\
  $^{1}$University of Luxembourg, Luxembourg \\
  $^{2}$King Mongkut's University of Technology Thonburi, Thailand \\
  \texttt{\{poomphob.suwannapichat,pascal.bouvry\}@uni.lu} \\
  \texttt{caesar.wu@ext.uni.lu} \\
  \texttt{boonyarit.chang@kmutt.ac.th}
}

\begin{document}
\maketitle
\begin{abstract}
LLM-based Multi-Agent Systems (MAS) achieve strong performance on complex reasoning tasks by coordinating multiple agents, but at the cost of substantial token consumption. Recent work on automatic topology design, ARG-Designer, has reframed this problem as autoregressive graph generation. However, its training objective provides no explicit incentive for the model to generate sparse and efficient topologies. We address this limitation by introducing a Reward-Guided Autoregressive Graph
Generation (RGA-Designer) inspired by Reinforcement Learning from Human Feedback (RLHF). We train a reward model that jointly captures task correctness and structural compactness, and then fine-tune the pretrained graph generator using the reward model as feedback. Our method preserves task accuracy at the level of ARG-Designer while reducing token consumption by an average of $20.5\%$.
\end{abstract}

\section{Introduction}



Large Language Models (LLMs) have recently addressed many problems that were considered challenging in the Natural Language Processing (NLP) domain. 
By framing other tasks in natural language, LLMs can be adapted to a wide array of applications, frequently achieving surprisingly strong performance.
Nevertheless, LLMs still make mistakes, particularly in complex tasks such as reasoning \cite{llms_fail_to_reason}. The reasoning capabilities of LLMs remain a subject of active debate. Because these models are fundamentally optimized for next-token prediction, it is difficult to claim that they truly reason rather than verbosely produce sequences of tokens that are likely to follow a given context~\cite{transformers_cant_reason}. Several workarounds have been proposed to mitigate this issue, including chain-of-thought prompting~\cite{chain_of_thought} and self-verification~\cite{llm_self_verification}. Although these techniques do not address the root cause of the reasoning limitation, they have been shown to substantially extend the practical performance of LLMs on complex tasks.

LLM-based Multi-Agent Systems (MAS) represent another such workaround, in which multiple LLMs collaborate to solve complex problems. By assigning each LLM a specialized role, such as planner, coder, or critic, and orchestrating their interactions, MAS consistently achieve stronger performance than single-LLM baselines. 
However, MAS also increases inference cost as a trade-off for the improved performance. In static MAS, even simple tasks are processed through the same multi-step pipeline as complex ones, despite often being solvable with a single LLM inference. 
To make MAS architecture more dynamic, many existing works start from a predefined communication topology and either prune less important components or apply modifications to optimize it for each query~\cite{AgentPrune,agentdropout,gdesigner}. ARG-Designer~\cite{ARG_designer} adopts a different perspective, employing an autoregressive graph generator that constructs topologies from scratch. This design enables the topologies to be more flexible and is not constrained by predefined templates. However, ARG-Designer trains the graph generator by maximizing the likelihood of topologies seen in the training set; under this objective, the model has no explicit incentive to favour sparser, more efficient structures.


We incorporate a reward-guided training scheme inspired by Reinforcement Learning from Human Feedback
(RLHF)~\cite{rlhf} which guides the graph generator toward higher-quality graphs through a learned reward model that accounts for both graph size and task correctness.
\footnote{Code and detailed hyperparameter settings of the experiments are available at \url{https://github.com/psuwannapich/RGA-Designer}.} 
With this method, the graph generator is no longer restricted to reproducing structures seen during training; instead, it is encouraged to discover sparser topologies while preserving task performance. Our main contributions are as follows:
\begin{itemize}
    \item We apply reward-guided training schema to autoregressive MAS topology generation, addressing a limitation of likelihood-based objectives that have no incentive for structural compactness.
    \item We design a graph-level reward model that jointly captures task correctness and structural compactness.
    \item Across six benchmarks, our approach reduces token consumption by an average of $20.5\%$ over ARG-Designer while preserving task accuracy.
\end{itemize}

\section{Related Works}


This section first reviews LLM-based multi-agent systems and the evolution of topology design from static structures to learned, task-adaptive configurations. It then introduces Reinforcement Learning from Human Feedback (RLHF), the training paradigm adapted to guide the graph generator. Finally, it discusses Graph Neural Networks (GNNs), which serve as the backbone of the reward model.

\subsection{LLM-based Multi-Agent Systems}




LLM-based Multi-Agent Systems can be formalized as a directed acyclic graph (DAG), where nodes represent LLM agents with specific roles and edges represent information shared between them. The design of the collaboration graph and the selection of agent roles are crucial for the overall performance of a Multi-Agent System.

Early work on MAS has proposed a range of static topologies, including chains, stars, and debate-style configurations~\cite{self_refine_chain_mas,hong2024metagpt,llm_debate,chateval,GPTSwarm}. Other approaches draw inspiration from real-world workflows, such as role decomposition inspired by software development workflows~\cite{chatdev}. 

Subsequent works frame the MAS design problem as query-based topology adaptation. AgentPrune \cite{AgentPrune} uses a predefined template as the starting point then defines communication redundancy in MAS and performs pruning on graph edges, achieving performance comparable to dense baselines at a fraction of the inference cost. AgentDropout \cite{agentdropout} expands this idea by eliminating low-contribution agents (graph nodes) via adjacency matrix optimization. However, both methods start from a predefined template and learn which nodes or edges to remove. They cannot create unseen structures.
Rather than pruning predefined structures, G-Designer \cite{gdesigner} employs a variational graph autoencoder that encodes a template graph topology along with a task-specific virtual node then decodes a task-adaptive graph with sparsity regularization. Although G-Designer goes beyond pruning, it still optimizes within fixed template graphs. ARG-Designer \cite{ARG_designer}, the direct predecessor of our work, reframes MAS topology design as an autoregressive graph generation task.


The collaboration graph is generated from scratch by iteratively producing nodes and edges until a termination signal is reached. At step $i$, the model samples the role of the next agent $v_i$ conditioned on the partial graph $\mathcal{G}_{<i}$ before step $i$, the task query $q$, and the available role pool $\mathcal{R}$:

\begin{equation}
    \label{eq:node_gen}
    v_i \sim P(v_i \mid \mathcal{G}_{<i}, q, \mathcal{R}).
\end{equation}

If the sampled role corresponds to the special \texttt{END} token, generation terminates. Otherwise, $v_i$ is added to the graph, and the model then determines its incoming edges by sampling the existence of an edge from each previously generated node $v_j$ ($j < i$):

\begin{equation}
    \label{eq:edge_gen}
    e_{j,i} \sim P(e_{j,i} \mid v_i, \mathcal{G}_{<i}, q).
\end{equation}

The graph generation model is trained using a supervised paradigm: candidate graph $\mathcal{G}_i$ is executed against its training query $q_i$ and only the graph and query pairs that can complete the task correctly will be included in the dataset $\mathcal{D}$. The graph generation model was trained on the dataset $\mathcal{D}$ in order to maximize the conditional log-likelihood of the ground-truth graphs given the query:

\begin{equation}
    \label{arg_loss}
    \mathcal{L}(\theta)= - \sum_{(\mathcal{G}, q) \in \mathcal{D}} \mathrm{log} P_\theta(\mathcal{G}|q)
\end{equation}

To ensure the sparsity of graph generation, the ARG-Designer training pipeline is separated into two steps: cold start and efficiency fine-tuning. Cold start step lets the model explore diverse topologies by creating graph/query pairs candidates using various well-known structures such as complete graphs and star graphs.
The efficiency fine-tuning step focuses on creating sparse but efficient graphs by including pruned graphs and verified efficient configurations.


\subsection{Reinforcement Learning from Human Feedback}


Reinforcement learning (RL) is a learning paradigm in which a model improves its policy by interacting with an environment and receiving rewards for the actions it takes. Historically, RL has been mostly applied in robotics and control, where an agent performs actions in a physical environment, observes the resulting reward signal, and adjusts its policy to maximize long-term return. Recently, RL has driven one of the most significant breakthroughs in language modeling: the development of ChatGPT. Beyond pre-training on a large text corpus for next-token prediction, ChatGPT incorporates human preferences to align its outputs with what users find helpful and appropriate. Human annotators are presented with pairs of candidate responses and asked to indicate which one is preferred. The resulting preference dataset is used to train a reward model, which learns to assign a score to a given response that is consistent with human judgment. The LLM is then fine-tuned via reinforcement learning, using the reward model as a proxy for human feedback to refine its output policy. This procedure is known as Reinforcement Learning from Human Feedback (RLHF)~\cite{rlhf}.

\subsection{Graph Neural Networks}



Graphs are widely used to represent relational data in many applications, including social network analysis, knowledge graphs, and information exchange in LLM-based Multi-Agent Systems. However, due to the nature of graph data, mapping graphs into numerical representations suitable for downstream analysis remains a challenging problem. The process of producing such numerical representations is referred to as graph embedding.
Early graph embedding methods adapted objectives from word embedding in the natural language processing (NLP) domain. An embedding model is trained to predict masked nodes in node sequences obtained by traversing the graph through random walks~\cite{deepwalk,node2vec}. However, these approaches cannot generalize to unseen nodes without re-running the embedding procedure on the new graph.


To enable graph embedding models to generalize to unseen graph structures without retraining, the framework of Message Passing Neural Networks (MPNNs) was introduced~\cite{message_passing_nn}. Each node is initialized with a feature vector, which can either be drawn at random or derived from node metadata. At each layer, every node sends a message to its direct neighbours; the receiving node then aggregates the incoming messages and uses them to update its own embedding. By stacking multiple MPNN layers, information can be propagated beyond direct neighbours to nodes that are several hops away. Building on this foundation, most modern graph embedding models are fundamentally constructed on the message-passing paradigm~\cite{gnn,gcn,graphSAGE}.


\section{Problem Definition}


We represent a multi-agent system (MAS) as a directed acyclic graph $\mathcal{G} = (\mathcal{V}, \mathcal{E})$ that captures the interactions between agents in the system. Each node $v_i \in \mathcal{V}$ corresponds to an LLM-based agent augmented with a predefined role prompt $R_i \in \mathcal{R}$, which specifies its function and expertise. A directed edge $(v_i, v_j) \in \mathcal{E}$ indicates that agent $v_i$ forwards its output $m_i$ as a textual message to agent $v_j$. To invoke agent $v_i$, the input prompt is assembled using template $f_T$ from three components: the role prompt $R_i$, the user query $q$, and the set of messages $\{m_j \mid (v_j, v_i) \in \mathcal{E}\}$ received from its direct predecessors, as formalized in Eq.~\ref{eq:agent_message}.

\begin{equation}
    \label{eq:agent_message}
    m_i=\mathrm{LLM}\Bigl( f_T(R_i, q,\{m_j|(v_j,v_i) \in \mathcal{E}  \}) \Bigr)
\end{equation}



To get the final answer of MAS, each agent $v_i \in \mathcal{V}$ is executed in topological order of graph $\mathcal{G}$ using Eq.~\ref{eq:agent_message} to produce its message $m_i$, and then aggregating the outputs of the terminal agents into a final answer:
\begin{equation}
    \label{eq:graph_execution}
    \hat{a} = \mathrm{Aggregate}\!\left(\{ m_i \mid v_i \in \mathcal{V}_{\mathrm{out}} \}\right),
\end{equation}
where $\mathcal{V}_{\mathrm{out}} \subseteq \mathcal{V}$ denotes the set of output agents. The task-completion indicator is then obtained by comparing $\hat{a}$ to the ground-truth answer $a^\star$:

\begin{equation}
    \label{eq:task_completion}
    c(\mathcal{G}, q) =
    \begin{cases}
        1, & \text{if } \hat{a} = a^\star, \\
        0, & \text{otherwise}.
    \end{cases}
\end{equation}

The goal of the MAS designer model $\pi_\theta$ is to generate a graph structure $\mathcal{G} \sim \pi_\theta(\mathcal{G} \mid q)$ from a given task query $q$ that successfully completes the task with the smallest possible structure as shown in following constrained objective:
\begin{equation}
    \label{eq:design_objective}
    \begin{split}
        \min_{\theta} \;\; \mathbb{E}_{q,\mathcal{G} \sim \pi_\theta(\mathcal{G} \mid q)} \Big[\, \lambda_\mathcal{V}\, |\mathcal{V}| \;+\; \lambda_\mathcal{E}\, |\mathcal{E}| \,\Big] \quad \\ \text{s.t.} \quad c(\mathcal{G}, q) = 1
    \end{split}
\end{equation}
where $|\mathcal{V}|$ and $|\mathcal{E}|$ are the number of agents and edges in $\mathcal{G}$, and $\lambda_\mathcal{V}, \lambda_\mathcal{E} \geq 0$ control the relative penalty on agent count and communication links.

\section{Method}
\label{sec:method}


We propose Reward-Guided Autoregressive Graph Generator (\textbf{RGA-Designer}), which applies the idea of RLHF to the graph generation model, guiding it to produce graphs that can complete the task while keeping the graph structure as compact as possible. Because graph quality is programmatically verifiable, we replace human feedback with explicit rule-based rewards that favor smaller, successful graphs. Our setup falls within the broader paradigm of \emph{Reinforcement Learning with Verifiable Rewards} (RLVR)~\cite{deepseek_grpo}, where an automated reward signal is provided in place of human annotation.

\begin{figure*}
    \includegraphics[width=\textwidth]{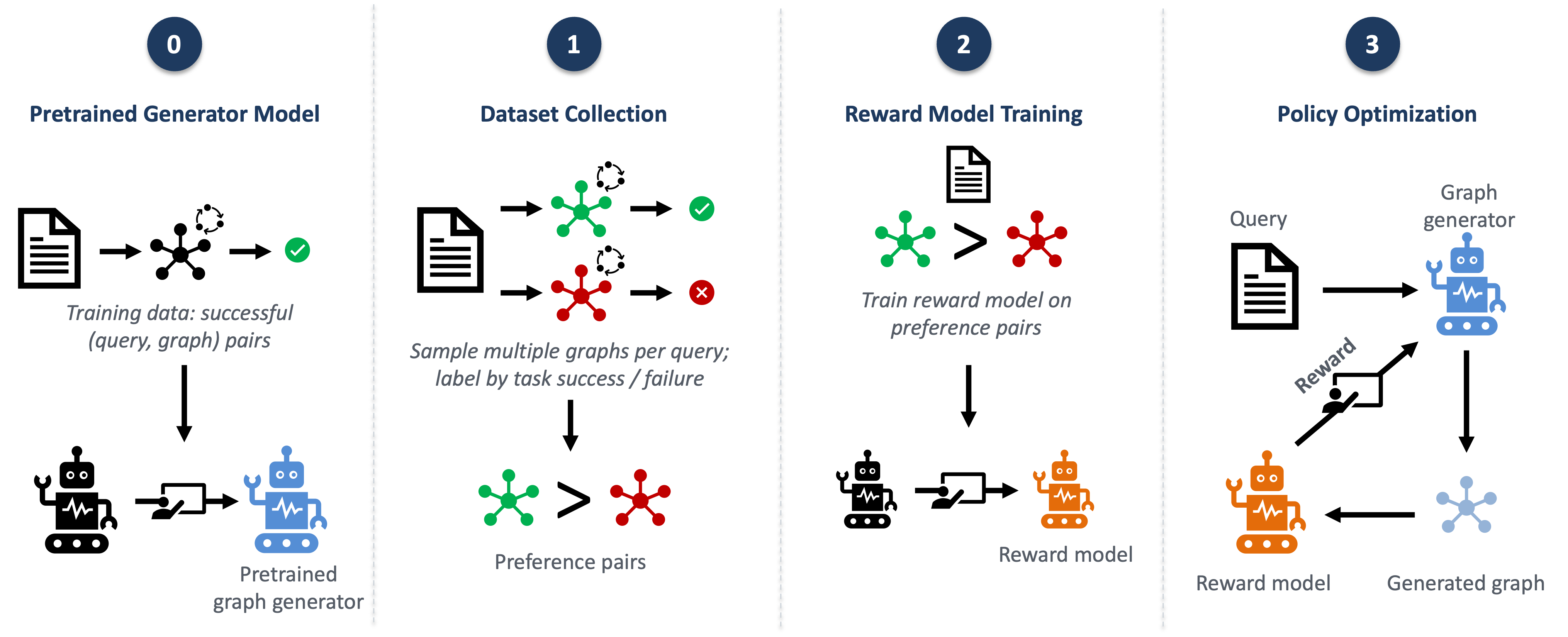}
    \caption{Overview of our reward-guided pipeline for MAS topology generation.}
    \label{fig:method_overview}
\end{figure*}

An overview of RGA-Designer is illustrated in Figure~\ref{fig:method_overview}. Our pipeline consists of three stages, preceded by a prerequisite pretraining stage inherited from ARG-Designer.
\textbf{(0) Pretrained Generator Model:}
following ARG-Designer, an autoregressive graph generator is trained on query/graph pairs.
\textbf{(1) Dataset Collection:}
for each task query, candidate graphs are sampled and labeled as successful or failed by executing them on the underlying MAS topology; the resulting samples are then organized into preference pairs.
\textbf{(2) Reward Model Training:}
a graph-level reward model is trained on the preference pairs, learning to assign higher scores to graphs that are both correct and compact.
\textbf{(3) Policy Optimization:}
the pretrained generator is fine-tuned with policy optimization, where the reward model guides the policy toward generating graphs with higher task success and lower structural complexity.


\subsection{Dataset collections}

We construct the dataset by executing LLMs over candidate graph structures and recording whether each structure successfully completes the task. However, we retain not only the samples that complete the task but also those that fail, since both are informative for training the reward model. Graphs are generated using the pretrained ARG-Designer model with varying sampling temperatures. Due to the autoregressive nature of ARG-Designer, multiple distinct graphs $\{\mathcal{G}^{(n)}_1, \mathcal{G}^{(n)}_2, \ldots, \mathcal{G}^{(n)}_M\}$ can be sampled from a single query $q^{(n)}$. To further reduce the data collection cost, we additionally include the training samples from both the cold-start and fine-tuning stages of ARG-Designer; since these graph/query pairs have already been executed during ARG-Designer training, incorporating them into our dataset incurs no additional cost.

Each training sample $s^{(n)}_i$ for the reward model is represented as a triplet $(\mathcal{G}^{(n)}_i, q^{(n)}, r^{(n)}_i)$. Every graph $\mathcal{G}^{(n)}_i = (\mathcal{V}^{(n)}_i,\mathcal{E}^{(n)}_i)$ is executed on the agentic system with query $q^{(n)}$, and its output is compared against the ground truth, yielding a binary completion flag $c^{(n)}_i = c(\mathcal{G}^{(n)}_i,q^{(n)})$. The ground-truth reward for the sample is then computed as:
\begin{equation}
    \label{eq:graph_reward}
    \begin{split}
    r^{(n)}_i = {} & \lambda_c c^{(n)}_i
      + \lambda_{\mathcal{V}} \max\left(0,
        \frac{\mathcal{V}_{\mathrm{max}} - |\mathcal{V}^{(n)}_i|}
             {\mathcal{V}_{\mathrm{max}} - 1}\right) \\
      & + \lambda_{\mathcal{E}} \operatorname{clip}_{[0,1]}\left(
        \frac{\mathcal{E}_{\mathrm{max}} - |\mathcal{E}^{(n)}_i|}
             {\mathcal{E}_{\mathrm{max}} - \mathcal{E}_{\mathrm{min}}}\right)
    \end{split}
\end{equation}
where $\lambda_c$, $\lambda_\mathcal{V}$, and $\lambda_\mathcal{E}$ are hyperparameters that weight task completeness, agent count, and edge count, respectively. $\mathcal{V}_{max}$ denotes the maximum number of agents allowed for each dataset (inherited from ARG-Designer), while $\mathcal{E}_{min}=|\mathcal{V}|-1$ and $\mathcal{E}_{max}=\frac{|\mathcal{V}| (|\mathcal{V}|-1)}{2}$ correspond to the minimum and maximum possible edge counts for a connected graph of size $|\mathcal{V}|$. 
The reward $r^{(n)}_i$ is designed to lie in the range $[0, 1]$, with the weights $\lambda_c$, $\lambda_\mathcal{V}$, and $\lambda_\mathcal{E}$ summing to $1$. The choice of weights reflects a  hierarchy in our design objective: task completion is treated as the dominant criterion ($\lambda_c = 0.6$), while structural compactness serves as a secondary preference ($\lambda_\mathcal{V} = 0.3$, $\lambda_\mathcal{E} = 0.1$). As a result, a candidate that fails to complete the task 
is penalized more than a candidate that succeeds but uses an inefficient topology, ensuring that correctness is never traded for sparsity.

To train the reward model within the RLHF framework, sample pairs must be constructed. Let $\mathcal{S}^{(n)}$ denote the set of training pairs associated with query $q^{(n)}$. The pairs in $\mathcal{S}^{(n)}$ are formed from all possible combinations, subject to two filtering conditions. First, at least one sample in each pair must successfully complete the task, as comparing two graphs that both fail provides little useful signal about graph quality. Second, the reward difference between the two samples must exceed a predefined margin $\delta$ (initially set to $0.05$):

\begin{equation}
\label{eq:training_pairs}
\mathcal{S}^{(n)} = \{ (s_i, s_j) \mid c_i = 1 \land r_i - r_j > \delta \}
\end{equation}

where the indices $i$ and $j$ range over the candidate samples collected for query $q^{(n)}$

\subsection{Reward Model}
\label{sec:reward_model}

The reward model $r_\theta$ is a graph neural network that scores a candidate graph $\mathcal{G} = (\mathcal{V}, \mathcal{E})$ conditioned on a task query $q$. To inject both task semantic and graph structure into the encoder, we construct each node feature vector $x_i$ as the concatenation of three components:

\begin{equation}
    \label{node_feature}
    x_i = \big[\, z_{r_i} \;\Vert\; z_q \;\Vert\; \phi_i \,\big]
\end{equation}
where $z_{r_i}$ is the role embedding of agent $v_i$, $z_q$ is the task query embedding.
Both embeddings are retrieved from \texttt{sentence-transformers/all-MiniLM-L6-v2} sentence encoder, yielding
$z_{r_i}, z_q \in \mathbb{R}^{384}$. $\phi_i \in \mathbb{R}^{5}$ is a vector of structural features comprising the graph size, edge count, generation-order position, in-degree, and out-degree. The graph is encoded by two GraphSAGE layers~\cite{graphSAGE} with residual connections and layer normalization. For each layer $\ell \in \{1, 2\}$:
\begin{equation}
    \label{eq:sage_layer}
    \begin{split}        
    h_i^{(\ell)} = \mathrm{ReLU}\!\Big(\mathrm{LN}\big(\mathrm{SAGE}^{(\ell)}(h^{(\ell-1)}, \mathcal{E})_i\big)\Big) \\ + W_{\mathrm{res}}^{(\ell)} h_i^{(\ell-1)}
    \end{split}
\end{equation}
where $h_i^{(0)} = x_i$, and $W_{\mathrm{res}}^{(\ell)}$ is a learnable projection for the residual. A graph-level embedding is then obtained by global mean pooling over all nodes, and a scalar reward is produced by a two-layer MLP head:
\begin{equation}
    \label{eq:reward_score}
    r_\theta(\mathcal{G}, q) = \mathrm{MLP}\!\left( \frac{1}{|\mathcal{V}|} \sum_{i \in \mathcal{V}} h_i^{(2)} \right)
\end{equation}

Given the preference pairs $\mathcal{S}^{(n)}$, the reward model is trained with the Bradley-Terry pairwise ranking objective~\cite{BradleyTerry}:



\begin{equation}
\label{eq:bt_loss}
    \begin{split}
    \mathcal{L}_{r}^{(n)}
      = -\,w_{c,r} \log \sigma\Big(
        r_\theta(\mathcal{G}^{(n)}_c, q^{(n)})
        \\ - r_\theta(\mathcal{G}^{(n)}_r, q^{(n)}) \Big)
    \end{split}
\end{equation}


where $\mathcal{G}^{(n)}_c$ and $\mathcal{G}^{(n)}_r$ denote the graphs of the chosen and rejected samples, respectively, and $w_{c,r} \in (0, 1]$ is a per-pair weight that down-scales pairs carrying less informative signal. In particular, pairs in which both samples successfully complete the task (\emph{pass/pass} pairs) convey only graph-size information, yet account for roughly 80--90\% of the training data. We therefore set $w_{c,r} = 1$ for pass/fail pairs and $w_{c,r} = \lambda_{p,p}$ for pass/pass pairs, where $\lambda_{p,p}$ is initially set to $0.1$.



An additional benefit of training a reward model is that its training data can be shared across datasets. Because each (query, graph) pair is encoded into a latent vector $x_i$ before being scored (Eq.~\ref{node_feature}), the reward model can seamlessly accommodate samples containing unseen agent roles. In contrast, ARG-Designer's node generator (Eq.~\ref{eq:node_gen}) explicitly conditions on the role pool $\mathcal{R}$ to score role candidates, meaning that any sample introducing a new role would require extending the node generator's classification head and retraining the model.

\subsection{Policy optimization}

Having trained reward model $r_\theta$, we then fine-tune graph generator policy $\pi_\theta$ to generate graphs that produce high reward $\hat{r}$ while
remaining close to the pre-trained ARG-Designer reference model $\pi_\mathrm{ref}$.
We adopt an on-policy variant of Group Relative Policy Optimization
(GRPO)~\cite{deepseek_grpo} for policy optimization.

Given the group of graph rewards $\{{\hat{r}^{(n)}_1, \ldots, \hat{r}^{(n)}_G}\}$ for query $q^{(n)}$, the advantage $\hat{A}^{(n)}_i$ for each candidate graph is computed by normalizing its reward as shown in Eq.~\ref{grpo_adventage}, where $\mu^{(n)}_r$ is the group mean, $\sigma^{(n)}_r$ is the group standard deviation. The denominator is lower-bounded by $\sigma_{\min} = 0.01$.


\begin{equation}
\label{grpo_adventage}
\hat{A}^{(n)}_i =
  \frac{\hat{r}^{(n)}_i - \mu^{(n)}_r}
       {\max\!\left(\sigma^{(n)}_r,\ \sigma_{\min}\right)}
\end{equation}



The graph generator policy $\pi_\theta$ is trained to maximize the advantage signal while constraining the generated distribution to remain close to the reference model $\pi_\mathrm{ref}$, as defined by the following objective:

\begin{equation}
    \label{eq:grpo_loss}
    \begin{aligned}
    \mathcal{L}_\pi(\mathcal{G}_i, q \mid \theta)
    = - \hat{A}_i \cdot \log \pi_\theta(\mathcal{G}_i \mid q)
      \\ + \beta \cdot \log \frac{\pi_\theta(\mathcal{G}_i \mid q)}{\pi_\mathrm{ref}(\mathcal{G}_i \mid q)}
    \end{aligned}
\end{equation}

where $\hat{A}_i$ is the group-relative advantage from Eq.~\ref{grpo_adventage}, and $\beta$ is a hyperparameter controlling the strength of the Kullback-Leibler (KL) regularization. 
Since we can draw a fresh group of $\mathcal{G}$ graphs from the current policy $\pi_\theta$ at every update step, the standard GRPO importance ratio $\rho_i$ equals $1$ when the gradient is calculated. Consequently, $\rho_i$ and the GRPO clipping term can be omitted.

\section{Experiments}


We evaluate our method on the same six benchmarks \cite{dataset_gsm8k,dataset_aqua,dataset_multiarith,dataset_swamp,dataset_mmlu,dataset_humaneval} used by ARG-Designer~\cite{ARG_designer}, summarized in Table~\ref{tab:datasets}. Each dataset was randomly divided into training, fine-tuning, and testing sets, consisting of 15, 25, and the remaining samples (capped at 500).

\begin{table}[htbp]
    \small
    \centering
    \caption{Summary of the six evaluation benchmarks used in our
    experiments. \#Test denotes the number of test instances sampled
    from each dataset for evaluation.}
    \label{tab:datasets}
    \renewcommand{\arraystretch}{1.2}
    \begin{tabular*}{\linewidth}{@{\extracolsep{\fill}}llcc@{}}
        \toprule
        \textbf{Category} & \textbf{Dataset} & \textbf{\#Test} & \textbf{Metric} \\
        \midrule
        Mathematical & GSM8K      & 500 & Accuracy \\
        reasoning    & AQuA       & 214 & Accuracy \\
                     & MultiArith & 500 & Accuracy \\
                     & SVAMP      & 500 & Accuracy \\
        \cmidrule{1-4}
        General reasoning & MMLU  & 500 & Accuracy \\
        \cmidrule{1-4}
        Code generation & HumanEval & 121 & Pass@1 \\
        \bottomrule
    \end{tabular*}
\end{table}


To ensure the stability of our results, every experiment is repeated over 10 independent runs, and we report the mean and standard deviation across runs. 
We benchmark our approach against five baselines: (1) \emph{Vanilla}, which relies on a single LLM call; (2) \emph{G-Designer}~\cite{gdesigner}; (3) \emph{AgentPrune}~\cite{AgentPrune}; (4) \emph{AgentDropout}~\cite{agentdropout}; and (5) \emph{ARG} or ARG-Designer ~\cite{ARG_designer}, the direct predecessor of our method.



Since MAS execution requires multiple LLM inferences and a long context window to support information exchange between agents, we adopt the open-source lightweight model \textbf{Qwen3-4B}~\cite{qwen_model} as the underlying LLM in this work. Qwen3-4B can be served on a single NVIDIA Tesla V100 GPU with 16~GB of VRAM handling long context lengths without running out of memory, making it well-suited for our multi-agent setting. We also disable Qwen3's \emph{thinking} mode to reduce the number of completion tokens generated per inference. 


As our reward model can be trained on graphs containing unseen roles (Section~\ref{sec:reward_model}), we train a single \emph{global} reward model on a combined preference dataset from all six benchmarks. This unified reward model is then used to supervise policy optimization across every benchmark. Furthermore, to ensure that the best topology is selected at inference time, we adopt a Best-of-$N$ (BoN) generation scheme: the policy samples $N$ (set $N = 5$ in our experiments) candidate graphs for each task query, and the candidate with the highest reward scored by the reward model is chosen as the final topology. Note that both the sampling and scoring steps are performed by the topology designer, not by an LLM. The overhead of BoN is negligible compared to LLM calls.
To assess whether our results differ significantly in terms of both accuracy and token usage, we apply Welch's $t$-test~\cite{Welch_t_test}, which tests for a difference in means between two distributions.

\begin{table*}[t]
\centering
\caption{Task accuracy (\%) across six benchmarks, reported as mean $\pm$ standard deviation over 10 runs.}
\label{tab:accuracy}
\small

\setlength{\tabcolsep}{3pt}
\begin{tabular}{lcccccc}
\toprule
Benchmark & Vanilla & G-Designer & AgentPrune & AgentDropout & ARG & RGA (Our) \\
\midrule
GSM8K      & 82.68$\pm$0.72 & 86.98$\pm$0.65 & 88.00$\pm$1.59 & \textbf{89.08$\pm$1.26} & 88.36$\pm$1.04 & 88.76$\pm$0.66 \\
AQuA       & 76.87$\pm$1.21 & 83.35$\pm$1.86 & 82.90$\pm$1.08 & \textbf{83.41$\pm$1.86} & 81.64$\pm$1.56 & 82.10$\pm$3.89 \\
MultiArith & 97.56$\pm$0.23 & 97.96$\pm$0.08 & 98.02$\pm$0.35 & 97.88$\pm$0.33 & \textbf{98.46$\pm$0.13} & 98.44$\pm$0.23 \\
SVAMP      & 90.20$\pm$0.27 & 94.28$\pm$0.21 & 94.80$\pm$0.84 & 94.54$\pm$0.38 & \textbf{94.98$\pm$0.85} & 94.52$\pm$0.49 \\
HumanEval  & 73.22$\pm$2.14 & 80.08$\pm$1.29 & 82.23$\pm$2.90 & 81.65$\pm$3.67 & 82.56$\pm$1.76 & \textbf{83.22$\pm$1.95} \\
MMLU       & 71.28$\pm$0.78 & 60.46$\pm$1.06 & 79.74$\pm$1.49 & \textbf{79.86$\pm$1.33} & 79.32$\pm$0.91 & 79.78$\pm$1.27 \\
\midrule
Average    & 81.97$\pm$0.56 & 83.85$\pm$0.52 & 87.61$\pm$0.68 & 87.74$\pm$0.64 & 87.55$\pm$0.49 & \textbf{87.80$\pm$0.60} \\
\bottomrule
\end{tabular}
\end{table*}

\begin{table*}[t]
\centering
\caption{Average tokens usage per task across six benchmarks, reported as mean $\pm$ standard deviation over 10 runs (rounded to integer). Best results among MAS methods (excluding the Vanilla baseline) are in \textbf{bold}.}
\label{tab:tokens}
\small

\setlength{\tabcolsep}{3pt}
\begin{tabular}{lcccccc}
\toprule
Benchmark & Vanilla & G-Designer & AgentPrune & AgentDropout & ARG & RGA (Our) \\
\midrule
GSM8K      & 224 $\pm$ 10 & 6104 $\pm$ 12 & 5694 $\pm$ 118 & 6447 $\pm$ 441 & 4546 $\pm$ 275 & \textbf{3863 $\pm$ 195} \\
AQuA       & 509 $\pm$ 97 & 9134 $\pm$ 67 & 5710 $\pm$ 263 & 5535 $\pm$ 266 & 3914 $\pm$ 162 & \textbf{2633 $\pm$ 191} \\
MultiArith & 135 $\pm$ 0 & 5347 $\pm$ 10 & 4997 $\pm$ 147 & 5644 $\pm$ 454 & 2824 $\pm$ 11 & \textbf{2754 $\pm$ 507} \\
SVAMP      & 142 $\pm$ 10 & 5242 $\pm$ 12 & 4921 $\pm$ 84 & 5748 $\pm$ 471 & 3816 $\pm$ 290 & \textbf{3349 $\pm$ 446} \\
HumanEval  & 262 $\pm$ 1 & 2851 $\pm$ 23 & 3043 $\pm$ 133 & 2659 $\pm$ 122 & 2238 $\pm$ 119 & \textbf{1715 $\pm$ 88} \\
MMLU       & 208 $\pm$ 16 & 10498 $\pm$ 47 & 11030 $\pm$ 1388 & 8353 $\pm$ 1353 & 5554 $\pm$ 798 & \textbf{3875 $\pm$ 538} \\
\midrule
Average    & 247 $\pm$ 18 & 6529 $\pm$ 13 & 5899 $\pm$ 258 & 5731 $\pm$ 251 & 3815 $\pm$ 124 & \textbf{3032 $\pm$ 205} \\
\bottomrule
\end{tabular}
\end{table*}



\begin{table*}[t]
\centering
\caption{Welch $t$-test $p$-values of RGA-Designer against each benchmark. \textbf{Bold} marks $p<0.05$. Arrows give the direction of the difference: for accuracy $\uparrow$ favours RGA-Designer, for token usage $\downarrow$ favours RGA-Designer.}
\label{tab:significance}
\small
\setlength{\tabcolsep}{3pt}
\begin{tabular}{lcccccc}
\toprule
Comparison & GSM8K & AQuA & MultiArith & SVAMP & HumanEval & MMLU \\
\midrule
\multicolumn{7}{l}{\textit{Accuracy}} \\
G-Designer             & \textbf{\textless\,0.001}$\uparrow$ & 0.376$\downarrow$ & \textbf{\textless\,0.001}$\uparrow$ & 0.180$\uparrow$ & \textbf{\textless\,0.001}$\uparrow$ & \textbf{\textless\,0.001}$\uparrow$ \\
AgentPrune   & 0.188$\uparrow$ & 0.544$\downarrow$ & \textbf{0.006}$\uparrow$ & 0.377$\downarrow$ & 0.384$\uparrow$ & 0.949$\uparrow$ \\
AgentDropout & 0.489$\downarrow$ & 0.354$\downarrow$ & \textbf{\textless\,0.001}$\uparrow$ & 0.920$\downarrow$ & 0.252$\uparrow$ & 0.892$\downarrow$ \\
ARG-Designer           & 0.320$\uparrow$ & 0.735$\uparrow$ & 0.814$\downarrow$ & 0.160$\downarrow$ & 0.437$\uparrow$ & 0.365$\uparrow$ \\
\midrule
\multicolumn{7}{l}{\textit{Token usage}} \\
G-Designer             & \textbf{\textless\,0.001}$\downarrow$ & \textbf{\textless\,0.001}$\downarrow$ & \textbf{\textless\,0.001}$\downarrow$ & \textbf{\textless\,0.001}$\downarrow$ & \textbf{\textless\,0.001}$\downarrow$ & \textbf{\textless\,0.001}$\downarrow$ \\
AgentPrune   & \textbf{\textless\,0.001}$\downarrow$ & \textbf{\textless\,0.001}$\downarrow$ & \textbf{\textless\,0.001}$\downarrow$ & \textbf{\textless\,0.001}$\downarrow$ & \textbf{\textless\,0.001}$\downarrow$ & \textbf{\textless\,0.001}$\downarrow$ \\
AgentDropout  & \textbf{\textless\,0.001}$\downarrow$ & \textbf{\textless\,0.001}$\downarrow$ & \textbf{\textless\,0.001}$\downarrow$ & \textbf{\textless\,0.001}$\downarrow$ & \textbf{\textless\,0.001}$\downarrow$ & \textbf{\textless\,0.001}$\downarrow$ \\
ARG-Designer           & \textbf{\textless\,0.001}$\downarrow$ & \textbf{\textless\,0.001}$\downarrow$ & 0.673$\downarrow$ & \textbf{0.014}$\downarrow$ & \textbf{\textless\,0.001}$\downarrow$ & \textbf{\textless\,0.001}$\downarrow$ \\
\bottomrule
\end{tabular}
\end{table*}


Table~\ref{tab:accuracy} and Table~\ref{tab:tokens} report task accuracy and token usage for each method, while Table~\ref{tab:significance} summarizes the $p$-values from Welch's $t$-test comparing our RGA-Designer against other baselines. In terms of accuracy, RGA-Designer yields minor improvements on several benchmarks; however, Welch's $t$-test indicates that most of these differences are statistically insignificant (whether improvements or degradations). According to Table~\ref{tab:significance}, none of the accuracy degradation is significant. In contrast, RGA-Designer delivers a substantial reduction in token usage, as clearly shown in Table~\ref{tab:tokens} and confirmed by the corresponding $p$-values in Table~\ref{tab:significance}. Token reduction is statistically significant on every benchmark except MultiArith when compared with ARG-Designer.


We further investigate why MultiArith is the only benchmark that does not yield a significant token reduction. 
Questions in MultiArith are generated from predefined templates, with new numerical values substituted. As a result, the questions exhibit limited linguistic variation and can be solved with simple topologies, leaving little structural redundancy for RGA-Designer to compress. In contrast, the other benchmarks contain more diverse natural-language questions whose solutions benefit from optimal collaboration structures, providing a larger margin for RGA-Designer to optimize the topology.


\begin{table*}[t]
\centering
\caption{Ablation study on three settings. \emph{Full} is our complete method; \emph{w/o RM} removes the global reward model; \emph{w/o BoN} removes Best-of-$N$ selection. \emph{w/o both} removes both components. Results are reported as mean $\pm$ standard deviation over 10 runs.}
\label{tab:ablation}
\small
\setlength{\tabcolsep}{2.1pt}
\begin{tabular}{l cc cc cc}
\toprule
& \multicolumn{2}{c}{\textbf{GSM8K}}
& \multicolumn{2}{c}{\textbf{HumanEval}}
& \multicolumn{2}{c}{\textbf{MMLU}} \\
\cmidrule(lr){2-3} \cmidrule(lr){4-5} \cmidrule(lr){6-7}
& Acc.\,$\uparrow$ & Tok.\,$\downarrow$
& Acc.\,$\uparrow$ & Tok.\,$\downarrow$
& Acc.\,$\uparrow$ & Tok.\,$\downarrow$ \\
\midrule
ARG-Designer
& 88.36 $\pm$ 1.04 & 4546 $\pm$ 275
& 82.56 $\pm$ 1.76 & 2238 $\pm$ 119
& 79.32 $\pm$ 0.91 & 5554 $\pm$ 798 \\
\midrule
Full
& \textbf{88.76 $\pm$ 0.66} & \textbf{3863 $\pm$ 195}
& \textbf{83.22 $\pm$ 1.95} & \textbf{1715 $\pm$ 88}
& 79.78 $\pm$ 1.27          & \textbf{3875 $\pm$ 538} \\
\quad w/o RM
& 88.72 $\pm$ 0.99 & 4166 $\pm$ 522
& 82.40 $\pm$ 2.17 & 1839 $\pm$ 116
& 79.42 $\pm$ 1.18 & 5288 $\pm$ 889 \\
\quad w/o BoN
& 88.28 $\pm$ 0.76 & 4281 $\pm$ 253
& 81.49 $\pm$ 2.45 & 1805 $\pm$ 119
& \textbf{79.84 $\pm$ 0.94} & 5056 $\pm$ 527 \\
\quad w/o both
& 88.72 $\pm$ 1.14 & 4207 $\pm$ 451
& 82.64 $\pm$ 2.02 & 1908 $\pm$ 95
& 78.36 $\pm$ 0.96 & 5425 $\pm$ 459 \\
\bottomrule
\end{tabular}
\end{table*}


Table~\ref{tab:ablation} reports the ablation study on three benchmarks, one from each task category: GSM8K (mathematical reasoning), HumanEval (code generation), and MMLU (general reasoning). The full method achieves the lowest token consumption. It also yields minor accuracy improvements on GSM8K and HumanEval, with the only exception being MMLU, where \emph{w/o BoN} achieves the highest accuracy. However, this slightly higher accuracy on MMLU comes at a substantial cost in token consumption which uses $30.5\%$ more tokens than the full method.

\section{Conclusion and Discussion}

In this work, we introduce RGA-Designer, a reward-guided training scheme that addresses a key limitation of ARG-Designer's original supervised objective that has no explicit incentive for the model to generate compact topologies. Building on the autoregressive graph generation paradigm, we employ a learned reward model to provide feedback to the graph generator, optimizing it to produce graphs that are sparse but still functional.

Across all benchmarks, RGA-Designer preserved task accuracy with no statistical differences, while achieving a substantial reduction in token consumption: on average, $20.5\%$ fewer tokens than ARG-Designer and reaching statistical significance on five of six benchmarks. The single exception, MultiArith, is a templated dataset that contains limited linguistic variation, leaving little redundancy for our method to compress.
Beyond the efficiency results, our framework offers an additional benefit. Because the reward model scores graphs in embedding space, its training data can be reused across datasets without architectural changes. In contrast, ARG-Designer's node generator conditions directly on the role pool $\mathcal{R}$, so introducing new agent roles requires extending the role classification head and retraining the generator.

\section*{Limitations}
\label{sec:limitations}



\paragraph{LLM backbone.}
Due to resource constraints, all experiments use Qwen3-4B as the underlying LLM, and the performance of RGA-Designer against baselines may shift when applied to different models. Our data construction pipeline depends on executing candidate graphs on a specific base LLM in order to label them as successful or failed. Switching to a different base model requires reconstructing the entire training dataset, and supporting heterogeneous MAS in which different agents are powered by different LLMs remains a challenge for our approach.

\paragraph{Reliance on ground-truth labels.}
While RGA-Designer can adapt topologies on a per-query basis, training the graph generator still requires ground truth. Therefore, our method is applicable to tasks with explicit verifiable answers (e.g., code generation, multiple-choice classification).
Extending the framework to open-ended tasks without verifiable ground truth is left to future work.

\section*{Acknowledgment}

We express our gratitude to Phatarapran Saraluck for providing valuable feedback for methodology design. The experiments presented in this paper were carried out using the HPC facilities of the University of Luxembourg~\cite{ulhpc}. 

\bibliography{custom}

@inproceedings{hong2024metagpt,
      title={Meta{GPT}: Meta Programming for A Multi-Agent Collaborative Framework},
      author={Sirui Hong and Mingchen Zhuge and Jonathan Chen and Xiawu Zheng and Yuheng Cheng and Jinlin Wang and Ceyao Zhang and Zili Wang and Steven Ka Shing Yau and Zijuan Lin and Liyang Zhou and Chenyu Ran and Lingfeng Xiao and Chenglin Wu and J{\"u}rgen Schmidhuber},
      booktitle={The Twelfth International Conference on Learning Representations},
      year={2024},
      url={https://openreview.net/forum?id=VtmBAGCN7o}
}

@inproceedings{llm_debate,
author = {Du, Yilun and Li, Shuang and Torralba, Antonio and Tenenbaum, Joshua B. and Mordatch, Igor},
title = {Improving factuality and reasoning in language models through multiagent debate},
year = {2024},
publisher = {JMLR.org},
booktitle = {Proceedings of the 41st International Conference on Machine Learning},
articleno = {467},
numpages = {31},
location = {Vienna, Austria},
series = {ICML'24}
}

@inproceedings{
    chateval,
    title={ChatEval: Towards Better {LLM}-based Evaluators through Multi-Agent Debate},
    author={Chi-Min Chan and Weize Chen and Yusheng Su and Jianxuan Yu and Wei Xue and Shanghang Zhang and Jie Fu and Zhiyuan Liu},
    booktitle={The Twelfth International Conference on Learning Representations},
    year={2024},
    url={https://openreview.net/forum?id=FQepisCUWu}
}

@inproceedings{chatdev,
    title = "{C}hat{D}ev: Communicative Agents for Software Development",
    author = "Qian, Chen  and
      Liu, Wei  and
      Liu, Hongzhang  and
      Chen, Nuo  and
      Dang, Yufan  and
      Li, Jiahao  and
      Yang, Cheng  and
      Chen, Weize  and
      Su, Yusheng  and
      Cong, Xin  and
      Xu, Juyuan  and
      Li, Dahai  and
      Liu, Zhiyuan  and
      Sun, Maosong",
    editor = "Ku, Lun-Wei  and
      Martins, Andre  and
      Srikumar, Vivek",
    booktitle = "Proceedings of the 62nd Annual Meeting of the Association for Computational Linguistics (Volume 1: Long Papers)",
    month = aug,
    year = "2024",
    address = "Bangkok, Thailand",
    publisher = "Association for Computational Linguistics",
    url = "https://aclanthology.org/2024.acl-long.810/",
    doi = "10.18653/v1/2024.acl-long.810",
    pages = "15174--15186"
}

@inproceedings{self_refine_chain_mas,
author = {Madaan, Aman and Tandon, Niket and Gupta, Prakhar and Hallinan, Skyler and Gao, Luyu and Wiegreffe, Sarah and Alon, Uri and Dziri, Nouha and Prabhumoye, Shrimai and Yang, Yiming and Gupta, Shashank and Majumder, Bodhisattwa Prasad and Hermann, Katherine and Welleck, Sean and Yazdanbakhsh, Amir and Clark, Peter},
title = {SELF-REFINE: iterative refinement with self-feedback},
year = {2023},
publisher = {Curran Associates Inc.},
address = {Red Hook, NY, USA},
booktitle = {Proceedings of the 37th International Conference on Neural Information Processing Systems},
articleno = {2019},
numpages = {61},
location = {New Orleans, LA, USA},
series = {NIPS '23}
}

@inproceedings{
    AgentPrune,
    title={Cut the Crap: An Economical Communication Pipeline for {LLM}-based Multi-Agent Systems},
    author={Guibin Zhang and Yanwei Yue and Zhixun Li and Sukwon Yun and Guancheng Wan and Kun Wang and Dawei Cheng and Jeffrey Xu Yu and Tianlong Chen},
    booktitle={The Thirteenth International Conference on Learning Representations},
    year={2025},
    url={https://openreview.net/forum?id=LkzuPorQ5L}
}

@inproceedings{agentdropout,
    title = "{A}gent{D}ropout: Dynamic Agent Elimination for Token-Efficient and High-Performance {LLM}-Based Multi-Agent Collaboration",
    author = "Wang, Zhexuan  and
      Wang, Yutong  and
      Liu, Xuebo  and
      Ding, Liang  and
      Zhang, Miao  and
      Liu, Jie  and
      Zhang, Min",
    editor = "Che, Wanxiang  and
      Nabende, Joyce  and
      Shutova, Ekaterina  and
      Pilehvar, Mohammad Taher",
    booktitle = "Proceedings of the 63rd Annual Meeting of the Association for Computational Linguistics (Volume 1: Long Papers)",
    month = jul,
    year = "2025",
    address = "Vienna, Austria",
    publisher = "Association for Computational Linguistics",
    url = "https://aclanthology.org/2025.acl-long.1170/",
    doi = "10.18653/v1/2025.acl-long.1170",
    pages = "24013--24035",
    ISBN = "979-8-89176-251-0"
}

@inproceedings{
    gdesigner,
    title={G-Designer: Architecting Multi-agent Communication Topologies via Graph Neural Networks},
    author={Guibin Zhang and Yanwei Yue and Xiangguo Sun and Guancheng Wan and Miao Yu and Junfeng Fang and Kun Wang and Tianlong Chen and Dawei Cheng},
    booktitle={Forty-second International Conference on Machine Learning},
    year={2025},
    url={https://openreview.net/forum?id=LpE54NUnmO}
}

@inproceedings{ARG_designer,
  title={Assemble your crew: Automatic multi-agent communication topology design via autoregressive graph generation},
  author={Li, Shiyuan and Liu, Yixin and Wen, Qingsong and Zhang, Chengqi and Pan, Shirui},
  booktitle={Proceedings of the AAAI Conference on Artificial Intelligence},
  year={2026}
}

@misc{rlhf,
      title={Training language models to follow instructions with human feedback}, 
      author={Long Ouyang and Jeff Wu and Xu Jiang and Diogo Almeida and Carroll L. Wainwright and Pamela Mishkin and Chong Zhang and Sandhini Agarwal and Katarina Slama and Alex Ray and John Schulman and Jacob Hilton and Fraser Kelton and Luke Miller and Maddie Simens and Amanda Askell and Peter Welinder and Paul Christiano and Jan Leike and Ryan Lowe},
      year={2022},
      eprint={2203.02155},
      archivePrefix={arXiv},
      primaryClass={cs.CL},
      url={https://arxiv.org/abs/2203.02155}, 
}

@inproceedings{deepwalk, series={KDD ’14},
   title={DeepWalk: online learning of social representations},
   url={http://dx.doi.org/10.1145/2623330.2623732},
   DOI={10.1145/2623330.2623732},
   booktitle={Proceedings of the 20th ACM SIGKDD international conference on Knowledge discovery and data mining},
   publisher={ACM},
   author={Perozzi, Bryan and Al-Rfou, Rami and Skiena, Steven},
   year={2014},
   month=Aug, pages={701–710},
   collection={KDD ’14} }

@inproceedings{node2vec,
author = {Grover, Aditya and Leskovec, Jure},
title = {node2vec: Scalable Feature Learning for Networks},
year = {2016},
isbn = {9781450342322},
publisher = {Association for Computing Machinery},
address = {New York, NY, USA},
url = {https://doi.org/10.1145/2939672.2939754},
doi = {10.1145/2939672.2939754},
booktitle = {Proceedings of the 22nd ACM SIGKDD International Conference on Knowledge Discovery and Data Mining},
pages = {855–864},
numpages = {10},
location = {San Francisco, California, USA},
series = {KDD '16}
}

@ARTICLE{gnn,
  author={Scarselli, Franco and Gori, Marco and Tsoi, Ah Chung and Hagenbuchner, Markus and Monfardini, Gabriele},
  journal={IEEE Transactions on Neural Networks}, 
  title={The Graph Neural Network Model}, 
  year={2009},
  volume={20},
  number={1},
  pages={61-80},
  doi={10.1109/TNN.2008.2005605}}

@inproceedings{gcn,
    title={Semi-Supervised Classification with Graph Convolutional Networks},
    author={Kipf, Thomas N. and Welling, Max},
    booktitle={International Conference on Learning Representations (ICLR)},
    year={2017}
}

@inproceedings{graphSAGE,
author = {Hamilton, William L. and Ying, Rex and Leskovec, Jure},
title = {Inductive representation learning on large graphs},
year = {2017},
isbn = {9781510860964},
publisher = {Curran Associates Inc.},
address = {Red Hook, NY, USA},
booktitle = {Proceedings of the 31st International Conference on Neural Information Processing Systems},
pages = {1025–1035},
numpages = {11},
location = {Long Beach, California, USA},
series = {NIPS'17}
}

@misc{deepseek_grpo,
      title={DeepSeekMath: Pushing the Limits of Mathematical Reasoning in Open Language Models}, 
      author={Zhihong Shao and Peiyi Wang and Qihao Zhu and Runxin Xu and Junxiao Song and Xiao Bi and Haowei Zhang and Mingchuan Zhang and Y. K. Li and Y. Wu and Daya Guo},
      year={2024},
      eprint={2402.03300},
      archivePrefix={arXiv},
      primaryClass={cs.CL},
      url={https://arxiv.org/abs/2402.03300}, 
}

@inproceedings{GPTSwarm, author = {Zhuge, Mingchen and Wang, Wenyi and Kirsch, Louis and Faccio, Francesco and Khizbullin, Dmitrii and Schmidhuber, J\"{u}rgen}, title = {GPTSwarm: language agents as optimizable graphs}, year = {2024}, publisher = {JMLR.org}, booktitle = {Proceedings of the 41st International Conference on Machine Learning}, articleno = {2597}, numpages = {25}, location = {Vienna, Austria}, series = {ICML'24} }

@article{dataset_gsm8k,
  title={Training Verifiers to Solve Math Word Problems},
  author={Cobbe, Karl and Kosaraju, Vineet and Bavarian, Mohammad and Chen, Mark and Jun, Heewoo and Kaiser, Lukasz and Plappert, Matthias and Tworek, Jerry and Hilton, Jacob and Nakano, Reiichiro and Hesse, Christopher and Schulman, John},
  journal={arXiv preprint arXiv:2110.14168},
  year={2021}
}

@inproceedings{
    dataset_mmlu,
    title={Measuring Massive Multitask Language Understanding},
    author={Dan Hendrycks and Collin Burns and Steven Basart and Andy Zou and Mantas Mazeika and Dawn Song and Jacob Steinhardt},
    booktitle={International Conference on Learning Representations},
    year={2021},
    url={https://openreview.net/forum?id=d7KBjmI3GmQ}
}

@inproceedings{dataset_multiarith,
    title = "Solving General Arithmetic Word Problems",
    author = "Roy, Subhro  and
      Roth, Dan",
    editor = "M{\`a}rquez, Llu{\'i}s  and
      Callison-Burch, Chris  and
      Su, Jian",
    booktitle = "Proceedings of the 2015 Conference on Empirical Methods in Natural Language Processing",
    month = sep,
    year = "2015",
    address = "Lisbon, Portugal",
    publisher = "Association for Computational Linguistics",
    url = "https://aclanthology.org/D15-1202/",
    doi = "10.18653/v1/D15-1202",
    pages = "1743--1752"
}

@inproceedings{dataset_swamp,
    title = "Are {NLP} Models really able to Solve Simple Math Word Problems?",
    author = "Patel, Arkil  and
      Bhattamishra, Satwik  and
      Goyal, Navin",
    editor = "Toutanova, Kristina  and
      Rumshisky, Anna  and
      Zettlemoyer, Luke  and
      Hakkani-Tur, Dilek  and
      Beltagy, Iz  and
      Bethard, Steven  and
      Cotterell, Ryan  and
      Chakraborty, Tanmoy  and
      Zhou, Yichao",
    booktitle = "Proceedings of the 2021 Conference of the North American Chapter of the Association for Computational Linguistics: Human Language Technologies",
    month = jun,
    year = "2021",
    address = "Online",
    publisher = "Association for Computational Linguistics",
    url = "https://aclanthology.org/2021.naacl-main.168/",
    doi = "10.18653/v1/2021.naacl-main.168",
    pages = "2080--2094"
}

@inproceedings{dataset_aqua,
    title = "Program Induction by Rationale Generation: Learning to Solve and Explain Algebraic Word Problems",
    author = "Ling, Wang  and
      Yogatama, Dani  and
      Dyer, Chris  and
      Blunsom, Phil",
    editor = "Barzilay, Regina  and
      Kan, Min-Yen",
    booktitle = "Proceedings of the 55th Annual Meeting of the Association for Computational Linguistics (Volume 1: Long Papers)",
    month = jul,
    year = "2017",
    address = "Vancouver, Canada",
    publisher = "Association for Computational Linguistics",
    url = "https://aclanthology.org/P17-1015/",
    doi = "10.18653/v1/P17-1015",
    pages = "158--167"
}

@misc{dataset_humaneval,
      title={Evaluating Large Language Models Trained on Code}, 
      author={Mark Chen and Jerry Tworek and Heewoo Jun and Qiming Yuan and Henrique Ponde de Oliveira Pinto and Jared Kaplan and Harri Edwards and Yuri Burda and Nicholas Joseph and Greg Brockman and Alex Ray and Raul Puri and Gretchen Krueger and Michael Petrov and Heidy Khlaaf and Girish Sastry and Pamela Mishkin and Brooke Chan and Scott Gray and Nick Ryder and Mikhail Pavlov and Alethea Power and Lukasz Kaiser and Mohammad Bavarian and Clemens Winter and Philippe Tillet and Felipe Petroski Such and Dave Cummings and Matthias Plappert and Fotios Chantzis and Elizabeth Barnes and Ariel Herbert-Voss and William Hebgen Guss and Alex Nichol and Alex Paino and Nikolas Tezak and Jie Tang and Igor Babuschkin and Suchir Balaji and Shantanu Jain and William Saunders and Christopher Hesse and Andrew N. Carr and Jan Leike and Josh Achiam and Vedant Misra and Evan Morikawa and Alec Radford and Matthew Knight and Miles Brundage and Mira Murati and Katie Mayer and Peter Welinder and Bob McGrew and Dario Amodei and Sam McCandlish and Ilya Sutskever and Wojciech Zaremba},
      year={2021},
      eprint={2107.03374},
      archivePrefix={arXiv},
      primaryClass={cs.LG},
      url={https://arxiv.org/abs/2107.03374}, 
}

@misc{qwen_model,
      title={Qwen3 Technical Report}, 
      author={An Yang and Anfeng Li and Baosong Yang and Beichen Zhang and Binyuan Hui and Bo Zheng and Bowen Yu and Chang Gao and Chengen Huang and Chenxu Lv and Chujie Zheng and Dayiheng Liu and Fan Zhou and Fei Huang and Feng Hu and Hao Ge and Haoran Wei and Huan Lin and Jialong Tang and Jian Yang and Jianhong Tu and Jianwei Zhang and Jianxin Yang and Jiaxi Yang and Jing Zhou and Jingren Zhou and Junyang Lin and Kai Dang and Keqin Bao and Kexin Yang and Le Yu and Lianghao Deng and Mei Li and Mingfeng Xue and Mingze Li and Pei Zhang and Peng Wang and Qin Zhu and Rui Men and Ruize Gao and Shixuan Liu and Shuang Luo and Tianhao Li and Tianyi Tang and Wenbiao Yin and Xingzhang Ren and Xinyu Wang and Xinyu Zhang and Xuancheng Ren and Yang Fan and Yang Su and Yichang Zhang and Yinger Zhang and Yu Wan and Yuqiong Liu and Zekun Wang and Zeyu Cui and Zhenru Zhang and Zhipeng Zhou and Zihan Qiu},
      year={2025},
      eprint={2505.09388},
      archivePrefix={arXiv},
      primaryClass={cs.CL},
      url={https://arxiv.org/abs/2505.09388}, 
}

@article{BradleyTerry,
 ISSN = {00063444, 14643510},
 URL = {http://www.jstor.org/stable/2334029},
 author = {Ralph Allan Bradley and Milton E. Terry},
 journal = {Biometrika},
 number = {3/4},
 pages = {324--345},
 publisher = {[Oxford University Press, Biometrika Trust]},
 title = {Rank Analysis of Incomplete Block Designs: I. The Method of Paired Comparisons},
 urldate = {2026-04-27},
 volume = {39},
 year = {1952}
}

@inproceedings{transformers_cant_reason,
author = {Dziri, Nouha and Lu, Ximing and Sclar, Melanie and Li, Xiang Lorraine and Jiang, Liwei and Lin, Bill Yuchen and West, Peter and Bhagavatula, Chandra and Le Bras, Ronan and Hwang, Jena D. and Sanyal, Soumya and Welleck, Sean and Ren, Xiang and Ettinger, Allyson and Harchaoui, Zaid and Choi, Yejin},
title = {Faith and fate: limits of transformers on compositionality},
year = {2023},
publisher = {Curran Associates Inc.},
address = {Red Hook, NY, USA},
booktitle = {Proceedings of the 37th International Conference on Neural Information Processing Systems},
articleno = {3081},
numpages = {40},
location = {New Orleans, LA, USA},
series = {NIPS '23}
}

@inproceedings{chain_of_Thought,
author = {Wei, Jason and Wang, Xuezhi and Schuurmans, Dale and Bosma, Maarten and Ichter, Brian and Xia, Fei and Chi, Ed H. and Le, Quoc V. and Zhou, Denny},
title = {Chain-of-thought prompting elicits reasoning in large language models},
year = {2022},
isbn = {9781713871088},
publisher = {Curran Associates Inc.},
address = {Red Hook, NY, USA},
booktitle = {Proceedings of the 36th International Conference on Neural Information Processing Systems},
articleno = {1800},
numpages = {14},
location = {New Orleans, LA, USA},
series = {NIPS '22}
}

@inproceedings{llm_self_verification,
    title = "Large Language Models are Better Reasoners with Self-Verification",
    author = "Weng, Yixuan  and
      Zhu, Minjun  and
      Xia, Fei  and
      Li, Bin  and
      He, Shizhu  and
      Liu, Shengping  and
      Sun, Bin  and
      Liu, Kang  and
      Zhao, Jun",
    editor = "Bouamor, Houda  and
      Pino, Juan  and
      Bali, Kalika",
    booktitle = "Findings of the Association for Computational Linguistics: EMNLP 2023",
    month = dec,
    year = "2023",
    address = "Singapore",
    publisher = "Association for Computational Linguistics",
    url = "https://aclanthology.org/2023.findings-emnlp.167/",
    doi = "10.18653/v1/2023.findings-emnlp.167",
    pages = "2550--2575"
}

@inproceedings{message_passing_nn,
author = {Gilmer, Justin and Schoenholz, Samuel S. and Riley, Patrick F. and Vinyals, Oriol and Dahl, George E.},
title = {Neural message passing for Quantum chemistry},
year = {2017},
publisher = {JMLR.org},
booktitle = {Proceedings of the 34th International Conference on Machine Learning - Volume 70},
pages = {1263–1272},
numpages = {10},
location = {Sydney, NSW, Australia},
series = {ICML'17}
}

@InProceedings{ulhpc,
    author = {S. Varrette and H. Cartiaux and S. Peter and E.
    Kieffer and T. Valette and A. Olloh},
    title = {{Management of an Academic HPC \& Research
    Computing Facility: The ULHPC Experience 2.0}},
    booktitle = {Proc. of the 6th ACM High Performance Computing and
    Cluster Technologies Conf. (HPCCT 2022)},
    year = {2022},
    month = {July},
    address = {Fuzhou, China},
    isbn = {978-1-4503-9664-6},
    publisher = {Association for Computing Machinery (ACM)},
}

@article{Welch_t_test,
    author = {Welch, B. L.},
    title = {THE GENERALIZATION OF ‘STUDENT'S’ PROBLEM WHEN SEVERAL DIFFERENT POPULATION variances ARE INVOLVED},
    journal = {Biometrika},
    volume = {34},
    number = {1-2},
    pages = {28-35},
    year = {1947},
    month = {01},
    issn = {0006-3444},
    doi = {10.1093/biomet/34.1-2.28},
    url = {https://doi.org/10.1093/biomet/34.1-2.28},
    eprint = {https://academic.oup.com/biomet/article-pdf/34/1-2/28/553093/34-1-2-28.pdf},
}

@misc{llms_fail_to_reason,
      title={Large Language Model Reasoning Failures}, 
      author={Peiyang Song and Pengrui Han and Noah Goodman},
      year={2026},
      eprint={2602.06176},
      archivePrefix={arXiv},
      primaryClass={cs.AI},
      url={https://arxiv.org/abs/2602.06176}, 
}

\appendix

%

\section{Hyperparameter Settings}
\label{app:hyperparameters}

Table~\ref{tab:hyperparams} consolidates hyperparameters used in our
experiments. All values were held fixed across the ten independent runs
reported in Section~5. $\mathcal{V}_{\mathrm{max}}$ is inherited from ARG-Designer and varies by benchmark: 4 for GSM8K, AQuA, MultiArith, and SVAMP, 5 for
HumanEval, and 6 for MMLU. The minimum number of agents is 2 for all
benchmarks.

\begin{table}[!h]
\small
\centering
\caption{Hyperparameter settings. Values are shared across all six
benchmarks unless noted otherwise.}
\label{tab:hyperparams}
\begin{tabular*}{\linewidth}{@{\extracolsep{\fill}}lll@{}}
\toprule
\textbf{Symbol} & \textbf{Description} & \textbf{Value} \\
\midrule
$\lambda_c$ & Task-completion weight & 0.6 \\
$\lambda_{\mathcal{V}}$ & Agent-count weight & 0.3 \\
$\lambda_{\mathcal{E}}$ & Edge-count weight & 0.1 \\
$\mathcal{V}_{\mathrm{max}}$ & Maximum agents allowed & 4 / 5 / 6 \\
$\mathcal{E}_{\mathrm{min}}$ & Minimum edges & $|\mathcal{V}|-1$ \\
$\mathcal{E}_{\mathrm{max}}$ & Maximum edges & $\tfrac{|\mathcal{V}|(|\mathcal{V}|-1)}{2}$ \\
$\delta$ & Minimum reward gap per pair & 0.05 \\
$z_{r_i}, z_q$ & Role / query embedding dim. & 384 \\
$\phi_i$ & Structural feature dim. & 5 \\
$w_{c,r}$ & Pair weight (pass/fail) & 1.0 \\
$\lambda_{p,p}$ & Pair weight (pass/pass) & 0.1 \\
$G$ & GRPO group size & 4 \\
$\sigma_{\min}$ & Advantage denominator floor & 0.01 \\
$\beta$ & KL regularization strength & 0.2 \\
$N$ & Best-of-$N$ candidates & 5 \\
\bottomrule
\end{tabular*}
\end{table}

\end{document}